%s4.tex
%\MAINTITLE{Analysis of the Irregular Pulsations of AC Her}

%\input aamod.cmm
 \input cp-aa.tex
%\input psfig
%\refereelayout
\overfullrule=0pt

\long\def\jumpover#1{{}}

\def\hf{\hfill}
\newdimen\thicksize
\newdimen\thinsize
\thicksize=1.8pt
\thinsize=0.6pt
\def\th{\thinspace}

\def\qquad{\quad\quad}
\def\ngth{\negthinspace}

\def\hf{\hfill}

\def\frac#1#2{{\textstyle{ #1 \over #2}}}

\def\eg{{{\it e.g.}\ }}

\def\cf{{\it cf.\ }}

\def\ie{{{\it i.e.}\ }}

\def\at{{\rm\char'100}}
\def\ni{\noindent}

%   solar

%
\def\bF{{\bf F}}
\def\bG{{\bf G}}
\def\bX{{\bf X}}

\def\bZ{{\bf Z}}

\MAINTITLE{Analysis of the Irregular Pulsations of AC Her}

 \AUTHOR{
 Z. Koll\'ath\th@2\th@1
 \FOOTNOTE
 {e--mail address: kollath\at buda.konkoly.hu \hfill},\th
 J. R. Buchler\th@1
 \FOOTNOTE
 {e--mail address: buchler\at phys.ufl.edu\hfill},\th
 T. Serre\th @3
 \& J. Mattei\th@4
 }

 \INSTITUTE{
 @1 Physics Department, University of Florida, Gainesville, FL 32611
 @2 Konkoly Observatory, Budapest, Hungary
 @3 Observatoire de Paris, Meudon, France
 @4 AAVSO, Cambridge, MA 02138
 }

 \DATE{June 1997}

 \ABSTRACT{The AAVSO lightcurve data of the irregularly pulsating star AC
Herculis of the RV Tau class are analysed.  The lightcurve is shown to be
incompatible with a periodic, or even multiperiodic pulsation, even if
allowance is made for evolution.  Instead the best explanation is that the
irregularly alternating cycles are a manifestation of low dimensional chaos.
The lightcurve is found to be generated by a 3 or 4 dimensional dynamics -- 3
or 4 first order ODEs.  The (Lyapunov) fractal dimension of the underlying
dynamic attractor is computed to be $d_L\approx$ 2.2, smaller than the value of
$d_L\approx$ 3.1 found for R~Sct.
 }

\KEYWORDS{Stars: individual: AC Her --
          Stars: oscillations --
          Stars: variable: others --
          Methods: data analysis -- 
          Methods: numerical -- 
          Chaos 
          }

\THESAURUS{06(08.09.2; 08.15.1; 08.22.3; 03.13.2; 03.13.4; 02.03.1;)}

%\begtabfullwid
%\halign{#\hfill \cr
%\ni {\bf Astronomy \& Astrophysics}
%\ni {\bf to be submitted, june 1997} \cr
%}
%\endtab

\def\rahmen#1{}

\maketitle

\input psfig

\titlea{\ \ INTRODUCTION}

In this paper we analyze the AAVSO data set of the lightcurve of
AC~Herculis that shows irregular pulsations with large cycle to cycle
alternations in the lightcurve.  The paper is organized in parallel to the
analysis of R~Scuti (Buchler, Serre, Koll\'ath \& Mattei 1995 hereafter
BSKM, Buchler, Koll\'ath, Serre \& Mattei 1996, hereafter BKSM).  Despite
their membership in the same RV~Tau class the lightcurves of these two
stars are quite different.  AC~Her has a period of 35 days, about half
that of R~Sct, has a much lower pulsation amplitude, and is much less
irregular.  From a practical point of view, the available AAVSO data set
is shorter, there are more gaps and fewer points per cycle, and the
relative noise level is higher, all of which make the analysis harder. 

In \S2 we show that the pulsations are incompatible with periodic or
multiperiodic behavior, even when evolution is allowed for.  The pulsations
must therefore be either of a chaotic or of a stochastic nature.  (We follow
the general trend of distinguishing the two by calling chaotic an erratic
signal that is generated by a deterministic low dimensional dynamics whose form
can be found, at least in principle, and stochastic a very high dimensional
dynamics that is too complicated to be characterized otherwise than
stochastically.)  In \S3 we perform a nonlinear analysis that shows that the
observational data are indeed compatible with low dimensional chaos, and we
determine its quantitative properties.  This is followed by a discussion in
\S4, and we conclude in \S5.

 \begfig0cm
 % \vskip 6.2truecm
 \centerline{\psfig{figure=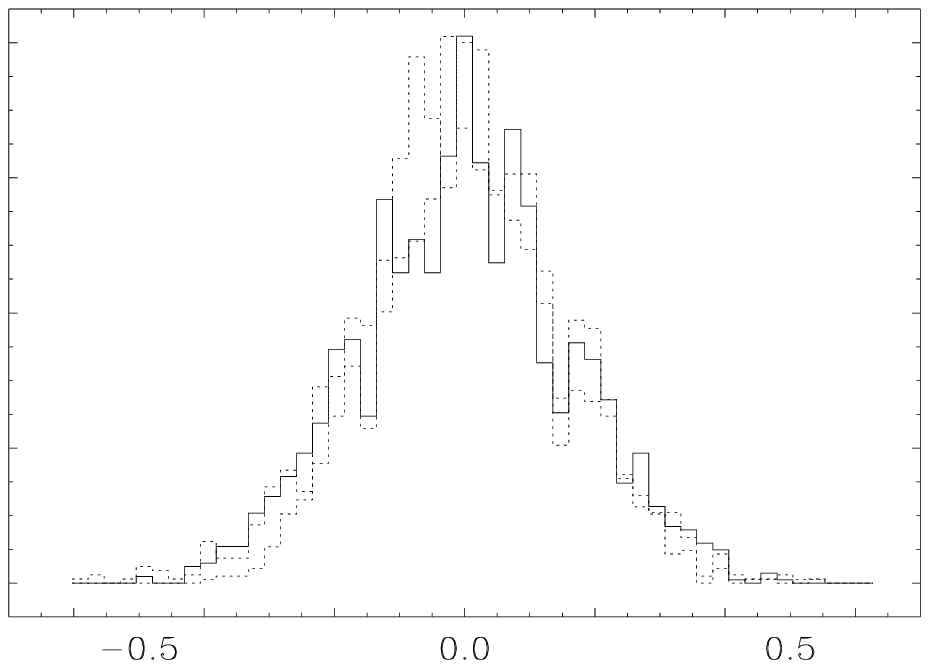,width=9.truecm}}
 \figure{1} {Error distributions of the AAVSO data; data grouped into 3 ranges,
below, in and above 7.5 to 7.8~mag.
 %\vskip -6.5truecm
 }
 \endfig

\titlea{\ \ Standard Analysis}

 \begfig0cm
   \centerline{\psfig{figure=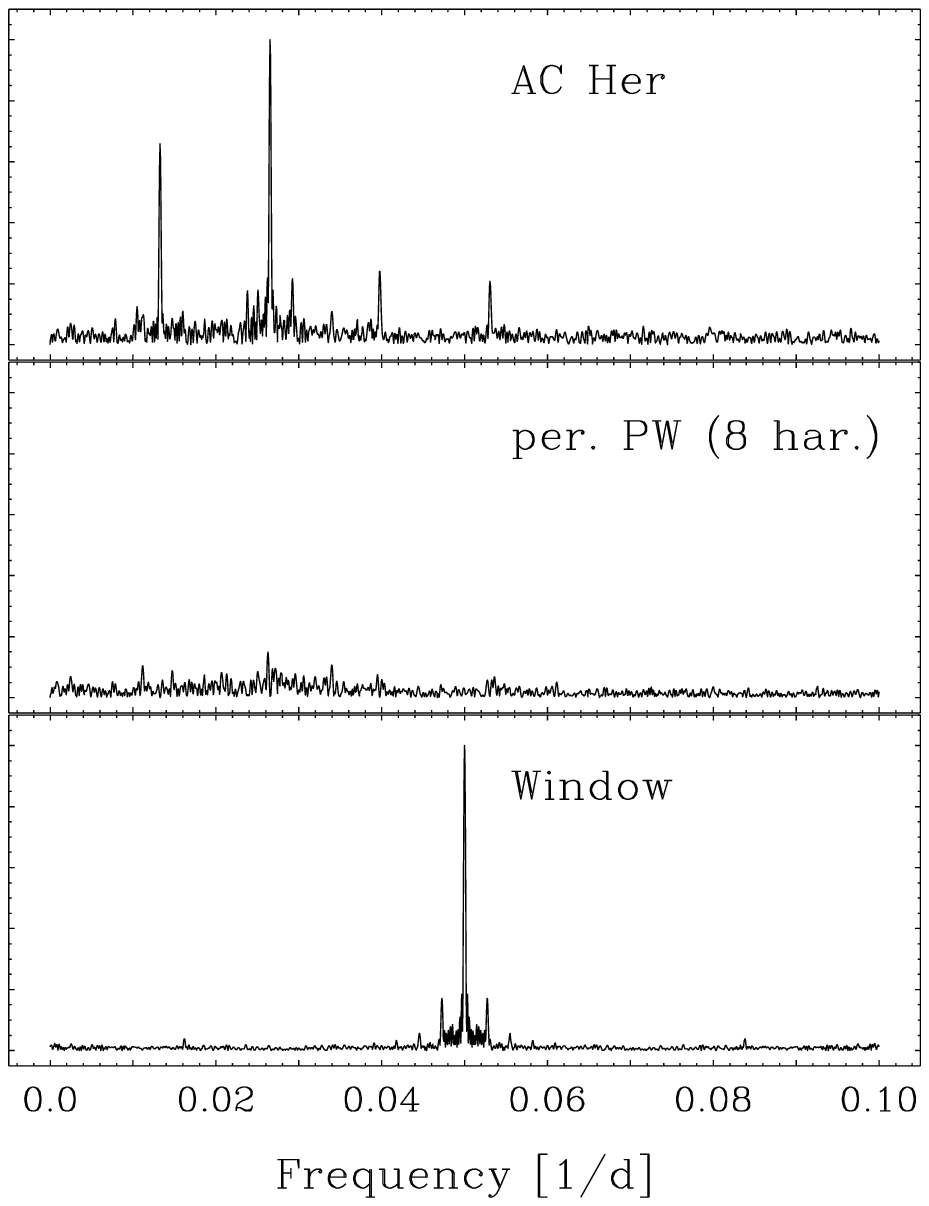,width=9.truecm}}
 \figure{2a} {
 Fourier Spectra. {\sl Top row:} AAVSO data;
 {\sl 2nd row:} periodic prewhitening with $f_0$ and 7 harmonics (same ordinate
scale);
 {\sl 3rd row:} spectral window.
 }
 \endfig

For a description of the AAVSO observational data archive we refer to
Percy and Mattei 1993. The data set of visual observations spans JD2437596
-- 2449442.  For most of our nonlinear analyses we use only the second
part (after JD2445000) of the data where the data coverage is better.  The
average (rms) error of the data is quite large, about 0.15~mag.  However,
the errors are found to have a Gaussian distribution, independent of {\sl
magnitude}, but not of luminosity, which is a reflection of the visual
nature of the observations.  This is quite apparent from Fig.~1 where we
have superposed the histograms for the errors for the data spit into three
groups, viz. below, in and above the range 7.5 -- 7.8~mag.  Just as for
R~Sct we therefore analyze the magnitude rather than the more physical
luminosity.  It is thanks to the Gaussian error distribution that in the
next section we are able to extract a relatively good average signal from
the bare data. 

In Fig.~2a, on top, we display the {\sl amplitude} Fourier spectrum (FS) of the
averaged AAVSO data for the observed AC~Her lightcurve.  The spectrum is
clearly dominated by four peaks.  The fundamental frequency peak is located at
$f_0$=0.01326~$d^{-1}$ \th ($P_0$=75.43~d).

 \begfig0cm
   \centerline{\psfig{figure=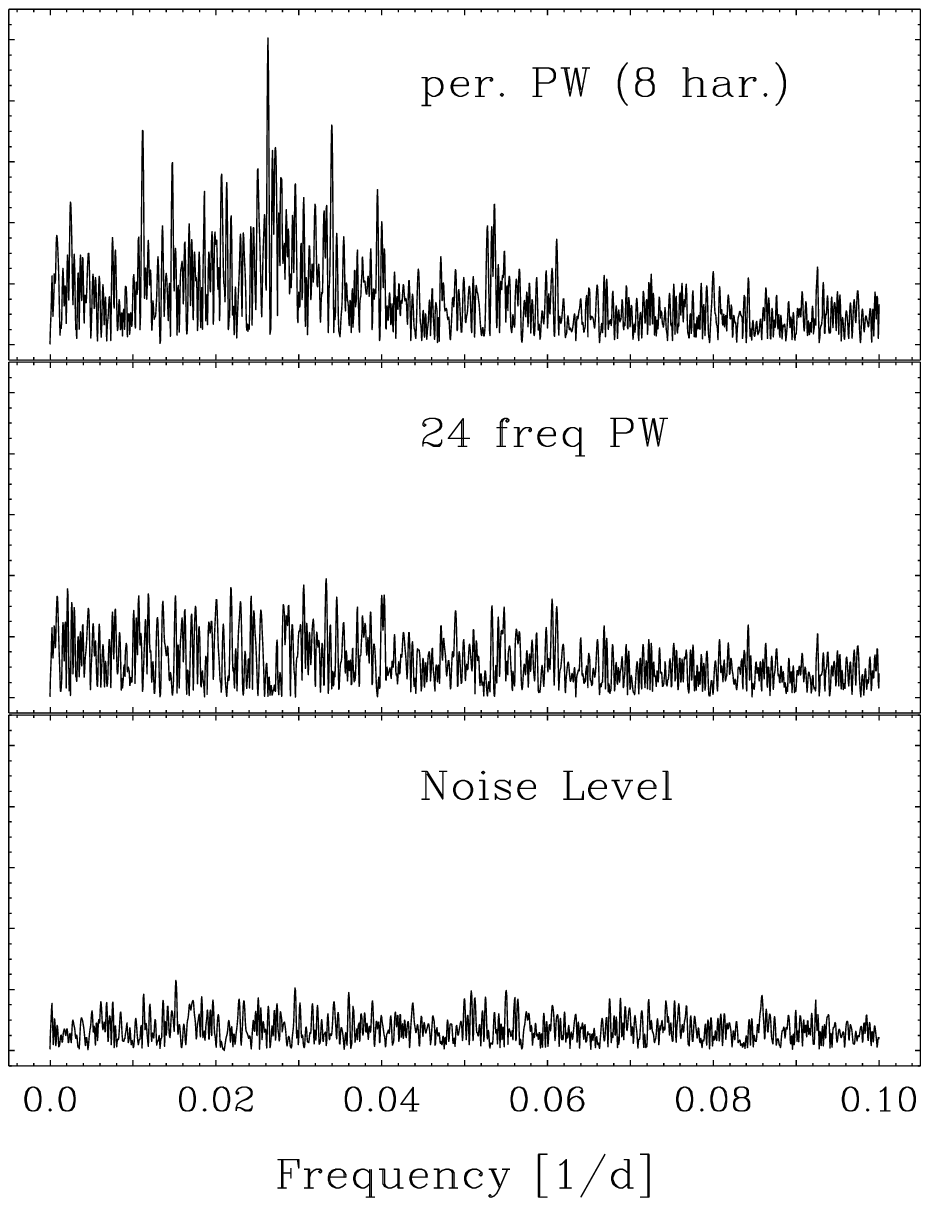,width=9.truecm}}
 \figure{2b} {
 Fourier Spectra;  {\sl Top row:} periodic prewhitening with $f_0$ and 7
harmonics;
 {\sl 2nd row:} multiperiodic prewhitening with $f_0$ and 7 harmonics plus
the 23 most important independent frequencies;
 {\sl bottom row:} expected observational noise level;
all three figures have the same ordinate scale.
 }
 \endfig

In the bottom row we show the spectral window which introduces some of the
side-lobe structure that is evident in the top figure.

Prima facie the FS of AC~Her seems to suggest periodicity (as does the
lack of phase jumps in the pulsations, cf. Fig.~5 below).  Therefore, in
the second row we show the FS of the lightcurve with a periodic
prewhitening with $f_0$ and 7 harmonics. \th\th (Actually the last,
highest four harmonics improve the prewhitening very little).  The phases
of the three dominant harmonics, $\Phi_k$, \th k=1, 3\th vary in
concordance with $\Phi_0$ over the time span of the data (see Fig.~3 where
we show $\Phi_0$ and $\Phi_1$).  This establishes a strong harmonic or
perhaps resonant connection among the four dominant peaks.  However, the
'grass' in the FS is very high, and furthermore the lightcurve has several
observationally well established local features that a periodic fit
completely misses. 

To investigate the possibly noisy origin of the left-over grass we take the
periodic component (that has been used in the prewhitening of the second row of
Fig.~2b) and add simulated Gaussian observational noise, with an intensity as
indicated by the AAVSO data, viz. $\approx$ 0.15~mag.  The FS of a prewhitened
realization of such a noisy periodic signal is shown in the bottom row of
Fig.~2b.  Note that the ordinate scales of figures 2a and 2b are different.
For comparison the same, periodically prewhitened FS is therefore shown in
both figures.

Could the pulsations be multiperiodic instead of periodic?  To answer this
question we first allow the eight dominant frequencies (i.e. $f_0$ and
previously its harmonics) to be independent.  This multiperiodic prewhitening
produces essentially no reduction in the grass of the FS compared to the
periodic prewhitening, nor does it improve the fit to the data.  Next we make a
more extensive prewhitening using now $f_0$ and its 7 harmonics plus 24 {\sl
linearly independent} frequencies, chosen from the highest peaks in row 2 of
Fig.~2a.  The resultant FS is shown in the second row FS of Fig.~2b.  Despite
this large number of frequencies quite a bit of unresolved grass still remains
in the FS that cannot be explained away as observational noise as a comparison
with the last row indicates.  In addition, an overlay of the 24--periodic fit
on the AAVSO data shows that many significant features are still missed.

Furthermore one would be hard-pressed theoretically to explain the origin of 24
or more frequencies of oscillation in this star, particularly if the star is
undergoing radial pulsations, as expected.  Even if nonradial pulsations are
involved and these visual observations could resolve up to $\ell$=2 modes,
it would not be possible to account for all these 'observed' frequencies.

 \begfig0cm
   \centerline{\psfig{figure=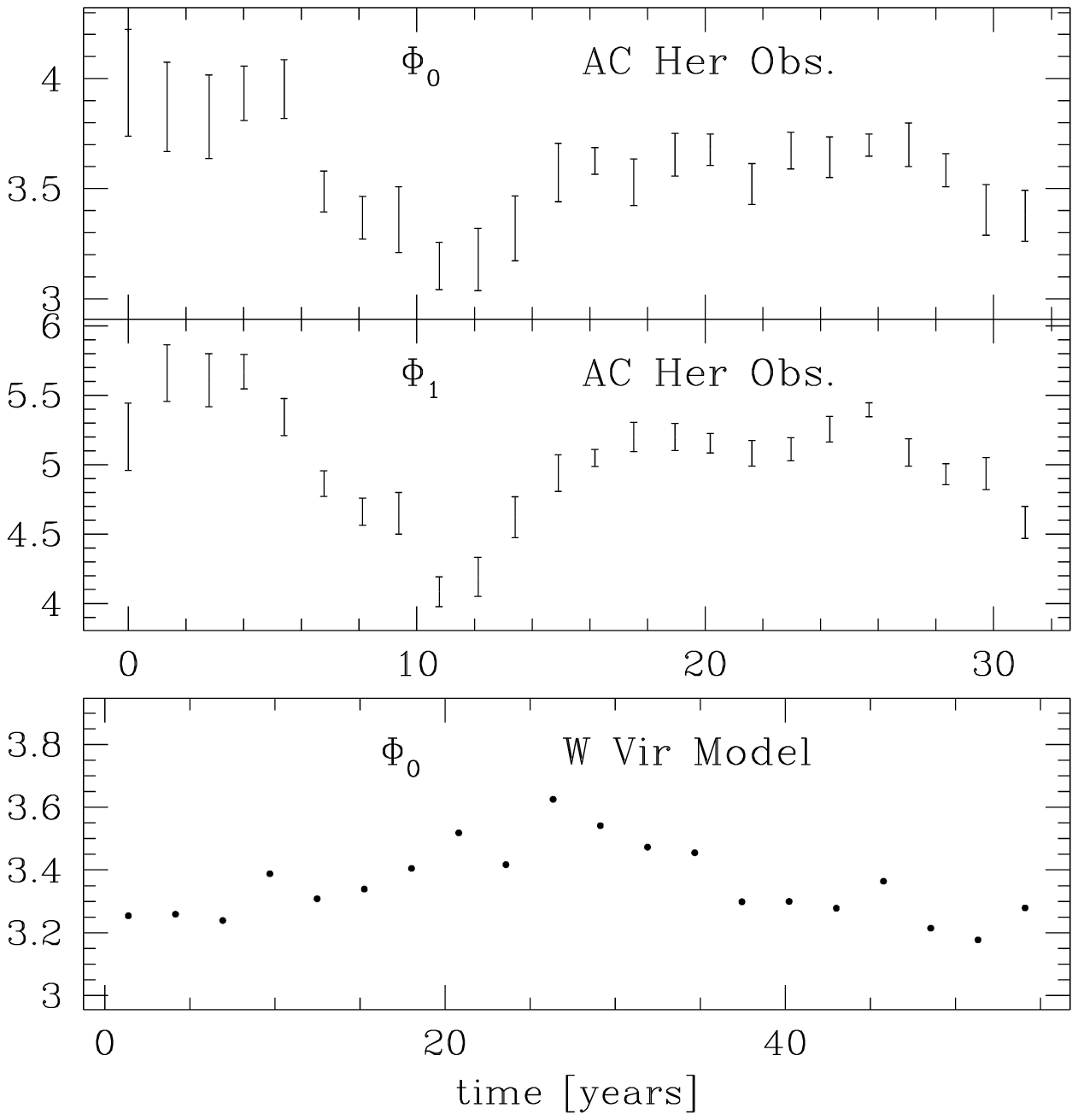,width=9.truecm}}
 \figure{3} {
 {\sl Top:} Time-dependence of the phases $\Phi_0$ and $\Phi_1$ over the span
of the AC~Her data set, 
 {\sl bottom:} Variation of the phase $\Phi_0$ with time over the span of a
corresponding data set of W~Vir numerical model pulsations.
 }
 \endfig

 \begfig0cm
   \centerline{\psfig{figure=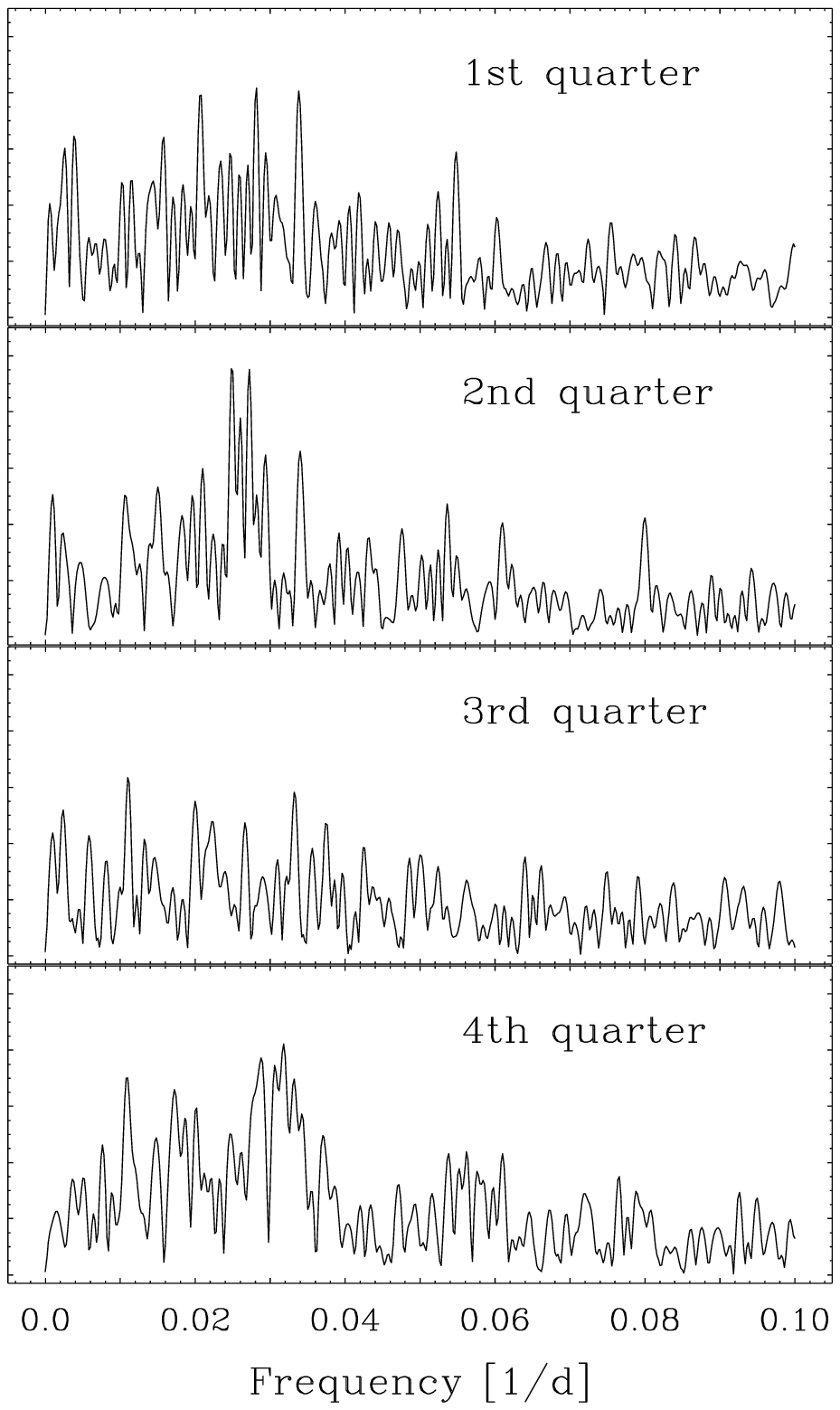,width=9.truecm}}
 \figure{4} {
 Fourier Spectra of the prewhitened successive quarter partitions of the AAVSO
data.
 }
 \endfig

The lightcurve therefore {\sl cannot reasonably be explained as being periodic
or multiperiodic with constant frequencies}.

Could the star be undergoing noticeable evolution over the decade that the data
span?  Fig.~3 reports the time-dependence of the phases $\Phi_0$ and $\Phi_1$
of the lightcurve for the frequencies $f_0$ and $f_1=2f_0$.  Here we 
have used the whole AAVSO dataset.  Were one to
interpret the phase variations as due to evolutionary changes in the frequency
one might conclude from the two approximately parabolic arcs that the frequency
changes linearly in time ($\dot P_0$ = 4$\times 10^{-4}$), but that in addition
there is a very strong frequency glitch around $t$=11~y.  However, one would be
hard pressed to come up with a mechanism for such a large glitch.
Zsoldos (1988) found a periodic variation in the $O-C$ curve of the
minima with a period of 9323 days. We cannot check the existence
of this modulation because of the shorter time span of the data.
Very recently Percy et al. (1977) have studied the period changes
of RV Tauri stars, including AC Her. They concluded that the $O-C$
diagrams can be interpreted as a superposition of random errors 
in the measured times of minimum, and random cycle-to-cycle 
fluctuations in the period.

 \begfigwid0cm
 \centerline{\psfig{figure=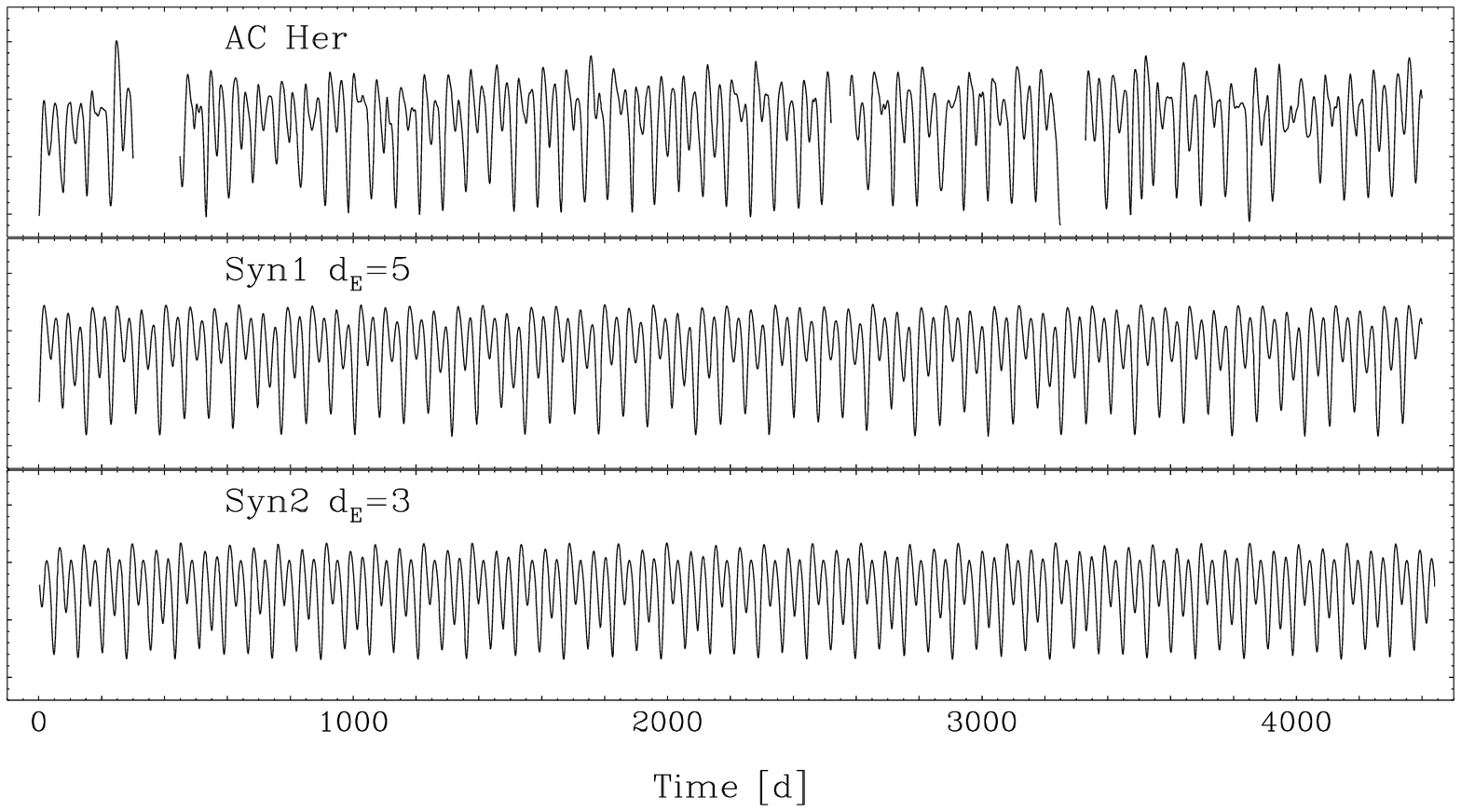,width=15.truecm}}
 \figure{5} {Lightcurves;
 {\sl Top}: AC Her, smoothed data (JD 2,445,000 + 4400~d);
 {\sl Middle:} section of synthetic data from reconstructed global map
 ($d_E$=5, $\Delta$=6, $p$=4, $\sigma$=0.065); {\sl Bottom:} synthetic
 data  ($d_E$=3, $\Delta$=5, $p$=6, $\sigma$=0.065) 
 }
 \endfig

If, despite the preceding discussion, one still insisted on interpreting the
pulsations as multiperiodic, then one should observe a systematic change in all
the frequencies and in the associated amplitudes.  We have therefore
partitioned the whole AAVSO data set into quarters.  Except for $f_0$ and its
three harmonics already considered there is no apparent systematic variation in
the grass.  Fig.~4 displays the FS of the successive partitions of the data,
all prewhitened and scaled the same way as the whole data set that has been
shown in the second row of Fig.~2b.

In order to offer an alternative explanation for the phase variations of
AC~Her, we also show in Fig.~3 the phase variations obtained from the magnitude
variations of hydrodynamic simulations of a W~Vir model (Kov\'acs \& Buchler
1986, Buchler 1993, Gouesbet et al. 1997).  Following the same reasoning as for
AC~Her one might interpret the phase variations again as indicating a linear
time dependence of the frequency (increasing with time here), also with an
abrupt frequency shift at 27~y.  However, here we are absolutely sure that such
an interpretation would be incorrect!  There is no evolution built into the
W~Vir model, and the pulsations were definitely shown to be chaotic.  The {\sl
phase variations are a natural result of the chaotic nature of the pulsations},
and over short time-intervals they can give an erroneous impression of
periodicity.  We therefore suggest that the apparent piecewise parabolic
behavior in AC~Her is spurious and that the phase variations have the same
chaotic origin.

We thus conclude that the pulsations of {\sl AC~Her cannot be explained as
those of an evolving multiperiodic star either}.

The presence of spots seems an unlikely explanation for the irregular
pulsations of this type of luminous star.  There is however the possibility of
stochasticity, for example caused by internal convection.  It is hard to rule
out a nonlinear stochastic process.  However, in the spirit of Occam's razor,
if a simpler valid explanation can be found we see no need to search for a
stochastic description which in any case would just push our ignorance further
down the line.

We are thus led to investigate if the pulsations can be due to an underlying
low dimensional chaotic dynamics.  In a priori support of such an explanation
we recall that the recent nonlinear analysis of the R~Sct lightcurve as well as
the modelling of W~Vir models have indicated such chaotic behavior.  We
therefore now turn to the type of nonlinear analysis that can uncover low
dimensional chaos when it is in the data.  (We refer a reader who is unfamiliar
with chaos to the excellent introduction by Ott 1993, or to general reviews by
Weigend \& Gershenfeld 1994, Abarbanel et al. 1993).

\titlea{\ \ Global Flow Reconstruction}

 In a recent publication Serre, Koll\'ath \& Buchler (1995, hereafter SKB; see
also Buchler 1997 for a lecture note, or Gouesbet et al. 1997 for more general
reviews) introduced a nonlinear time-series analysis, the global flow
reconstruction method, that is suitable for astronomical data.  A prior reading
of these papers is suggested, not only for a description of the method but also
for the notation and terminology that is used here.  The application to the
analysis of the lightcurve of R~Sct appears in BSKM and BKSM.

For the nonlinear analysis the AAVSO magnitude data first need to be converted
into a sequence with equal time-spacings, $\{s(t_n)\}$.  We do this by
performing 3~day averages that are then cubic-spline smoothed and interpolated
with a $\sigma$=0.065.\th For comparison we will also show results with a 5 day
average, and with smoothing parameters of $\sigma$=0.06 and $\sigma$=0.07.  The
reason for smoothing with a $\sigma$ that is less than the observational noise
is that we have already averaged the data which has introduced some smoothing.
The observational data have many gaps, many of which are too large to bridge.
We have therefore limited ourselves to the time-interval JD2445000 + 4400~days
which has only three gaps (cf. Fig.~5).

 \begfigwid0cm
 \centerline{\psfig{figure=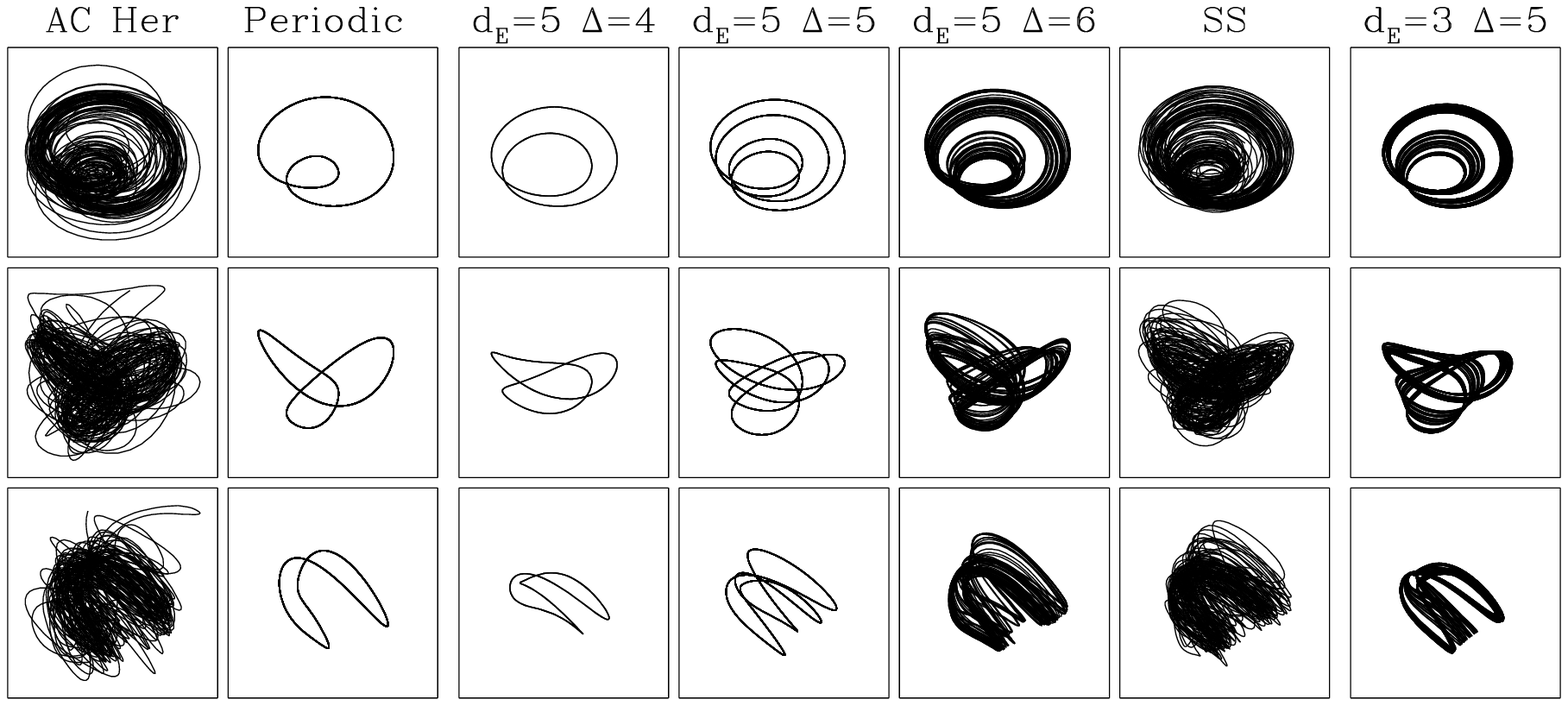,width=15.5truecm}}
 \figure{6} {Broomhead-King projections, from top down: ($\xi_2$ vs $\xi_1$),
 ($\xi_3$ vs $\xi_1$), ($\xi_3$ vs $\xi_2$); {\sl Col.~1:} smoothed AC~Her
lightcurve;
 {\sl Col.~2:} Periodic fit (8 harmonics) to the data;
 {\sl Cols.~3--5:} Synthetic signals for delay vectors,$d_E$=5, $\Delta$=4
through 6;
 {\sl Col.~6:} Synthethic signal of col.~ 5 ($\Delta$=6) with added noise,
sampled, averaged and smoothed like the AC~Her data (see text);
 {\sl Col.~7:} Synthetic signal for $d_E$=3, $\Delta$=5 
 }
 \endfig

Almost all nonlinear analyses start with the construction of 'delay
vectors'  ${\bX (t_n) \equiv  \bX}^n$
 $=\{s(t_n), s(t_n\ngth -\ngth\Delta), s(t_n\ngth -\ngth 2\Delta), \ldots ,
s(t_n\ngth -\ngth (d_E\ngth -\ngth 1)\Delta) \}$,
 where $\Delta$ is the 'delay' and $d_E$ is the dimension of the
'reconstruction space'.  The $\bX^n$ provide a strobed representation of the
'trajectory' of the system in this $d_E$--dim space.

We now make the a priori assumption that the light curve is generated by a
deterministic nonlinear mechanism of low dimension.  Phrased differently, we
assume that there is a dynamics that connects neighboring points of the
trajectory.  A theorem guarantees that there is a one-to-one correspondence
between the trajectory in the real or physical phase space of the system and
the trajectory in the reconstruction space.  Importantly, some of the
properties are preserved in this correspondence.  It is therefore possible to
extract information about the physical dynamics from the observational data of
a single observed variable (the luminosity here).  All we have to do is to
determine the functional form of the flow $d\bZ/dt = \bG(\bZ)$ that gives rise
to this trajectory $Z(t)$, where the observations provide the strobed values
$\bZ(t_n)=\bX^n$.

A priori we do not know what values to give to $\Delta$ and $d_E$.  The delay
should be large enough so that the attractor is not squashed into the diagonal,
but sufficiently small so that a polynomial nonlinearity can describe the flow.
There is a minimum value for $d_E$, called the embedding dimension, in which
the trajectory is resolved, i.e. it is devoid of intersections and cusps.
Nothing can be gained in increasing $d_E$ beyond this value, but as a check,
the properties of the reconstructed map should stay invariant in higher
dimensions.  We are of course interested in finding this minimum value because
it provides an upper bound on the dimensional of the physical dynamics that
underlies the observed signal (SKB).

 \begtabfull
 \tabcap{1} {Lyapunov exponents and dimension}
 \halign{
 \hf\th#\th\hf&
 \hf\th#\th\hf&
 \hf\th#\th\hf&
 \hf\th#\th\hf&
 \hf\th#\th\hf&
 \hf\th#\th\hf&
 \hf\th#\th\hf&
 \hf\th#\th\hf&
 \hf\th#\th\hf& 
 \hf\th#\th\hf\cr
 \noalign{\hrule\medskip}
$d_e$ & $\Delta$ & p & $\sigma$ & $\lambda_1$ & $\lambda_3$ &
 $\lambda_4$ & $\lambda_5$ & $d_{_{L}}$  \cr
 \noalign{\medskip\hrule\medskip}
 3 &   5 & 6 & 0.065 & 0.0033 & --0.034  &         &         & 2.10 
                                                           & $^\dagger$\cr
 3 &  10 & 6 & 0.065 & 0.0045 & --0.026  &         &         & 2.17 &\cr
 3 &  13 & 6 & 0.065 & 0.0069 & --0.030  &         &         & 2.23 &\cr
 4 &   5 & 5 & 0.065 & 0.0073 & --0.016  & --0.054 &         & 2.46 &\cr
 5 &   6 & 4 & 0.065 & 0.0045 & --0.023  & --0.025 & --0.032 & 2.19 
                                                           & $^\dagger$\cr
 5 &   7 & 4 & 0.065 & 0.0075 & --0.009  & --0.025 & --0.056 & 2.85 &\cr
 3 &  13 & 6 & 0.060 & 0.0015 & --0.029  &         &         & 2.05 &\cr 
 3 &   3 & 6 & 0.070 & 0.0021 & --0.037  &         &         & 2.06 &\cr
 3 &   6 & 6 & 0.070 & 0.0032 & --0.024  &         &         & 2.13 &\cr
 5 &   6 & 4 & 0.070 & 0.0043 & --0.015  & --0.025 & --0.036 & 2.29 &\cr
 3 &   5 & 6 & 0.065 & 0.0025 & --0.017  &         &         & 2.14 & $^*$ \cr
 3 &  10 & 6 & 0.065 & 0.0013 & --0.021  &         &         & 2.06 & $^*$ \cr
 3 &  13 & 6 & 0.065 & 0.0015 & --0.013  &         &         & 2.21 & $^*$ \cr
 \noalign {\medskip\hrule\medskip}
 \noalign {\ni Lyapunov exponents in d$^{-1}$}
 \noalign {\ni $^\dagger$ shown in Fig.~5}
 \noalign {\ni $^*$ with 5 day instead of 'standard' 3 day averages of AAVSO
data}
 }
 \endtab

For the quantitative reconstruction of the flow we turn to the global
reconstruction method (SKB).  We can equivalently reconstruct either a flow
$\bG$ or a 'map' $\bF$ that connects the neighboring points on the trajectory.
In SKB and BKSM we found that generally it is somewhat easier to construct
maps, and that is what we do here.  Our goal is thus to search for the best
global polynomial map $\bF$, such that $\bX^{n+1} = \bF(\bX^n)$.

Once we have constructed a map $\bF$ we can iterate it and produce 'synthetic'
signals that can then be compared to the observed lightcurve data.  Since both
are chaotic signals it would be meaningless to compare them point by point.
Instead they need to be compared for their overall properties (cf. SKB, BKSM).

In Fig.~5, rows~2 and 3, we show two of the best synthetic signals that we have
been able to construct in 5D and in 3D, respectively.  The first has the
parameters $\Delta$= 6, $p$= 4, $\sigma$=0.065, $d_E$= 5, the second,
$\Delta$= 5, $p$= 6, $\sigma$=0.065, $d_E$= 3.  (An earlier synthetic
signal was already shown in a review by Buchler, Koll\'ath \& Serre, 1995 where
we also show the first return maps).  The synthetic signals reproduce the data
reasonably well, although they are somewhat 'tamer'.

Broomhead-King projections which are projections on the eigenvectors of the
correlation matrix (q.v. SKB) provide a means to visualize the signals that is
optimal in many ways.  Thus in Fig.~6 we display the lowest Broomhead-King (BK)
projections for the smoothed AC~Her data in Col.~1.  From the top down we show
successively the projections $\xi_2$ vs.  $\xi_1$, $\xi_3$ vs. $\xi_1$, $\xi_3$
vs. $\xi_2$.  Col.~2 shows the 8~harmmonic periodic signal that has been used
to prewhiten the data earlier on (Fig.~2).  This periodic signal clearly
represents an average of the data.  The BK projections of the good synthetic
signals in 5D and 3D that have been shown in Fig.~5 are displayed in Cols.~5
and 7, respectively.  The signal of Col.~6 will be discussed below.

We need to pause to say a word about what we mean by 'successful' or 'best'
reconstructions.  The synthetic signals that we reconstruct with a map
generally are limit cycles, sometimes with period doubling (Col.~3),
quadrupling (Col.~4), or chaos.  We hardly ever encounter fixed point
attractors or multi-periodic ones (these could have 2 or 3 frequencies -- 2 or
at most 3--tori\th (since we keep $d_E\leq$ 6), except when the data are
oversmoothed (cf. e.g. BKSM).  Thus, the delay $\Delta$=3 gives a limit cycle,
whereas with $\Delta$=4 we get a 2-cycle (Col.~3 in Fig.~6), with $\Delta$=5 a
4-cycle (Col.~4) and with $\Delta$=6 a chaotic attractor (Col.~5).  We note
that the period two or four cycles that precede chaos when a parameter is
varied, such as $\Delta$ in the preceding example, generally indicate that the
parameters of the map are close to what it takes to give chaos, and instead of
being detrimental they are an indirect confirmation of chaos.

We have summarized some of the tests in Table~1.  The columns denote
successively the reconstruction dimension $d_E$ , the delay $\Delta$, the
highest order $p$ of the monomials used in the map, and the smoothing parameter
in the cubic spline fit to the AAVSO data.  Next are shown the Lyapunov
exponents of the synthetic signals $\lambda_k$, ordered with decreasing
magnitudes.  Finally, we show the fractal Lyapunov dimension $d_L$ which is an
overall measure of the chaos.
 
We draw attention to the following features: As expected all the
$\lambda_1>0$, which constitutes a clear indication of chaos.  Next, all
$\lambda_2\approx 0$ which because of the high sampling rate corroborates
that a flow (autonomous system) underlies the pulsations (cf. SKB).  The
synthetic signals have reasonably robust properties as we go to higher
dimensions $d_E$ = 4 and 5 (although for a reason not understood it is a
little harder to get good maps in $d_E$=4 than in 3 or 5).  The properties
are also reasonably independent of the 'massaging', i.e. averaging,
smoothing and interpolating of the observational data. 

From the Table it thus transpires that we are able to construct a good map in a
reconstruction space as low as $d_E$=3.  However, the map is not quite as good
as we can achieve with $d_E$=4 or 5.  It therefore cannot be ruled out that we
avoid all intersections and cusps in 3D \th (cf. SKB).

It is very comforting that the successful reconstructions are stable in the
sense that they give comparable properties in higher $d_E$ as Table~1 shows.

We do not display the FS of the synthetic signals, but note that whenever the
Lyapunov dimension is in the range 2.1 -- 2.2 the spectra compare rather well
with the observed spectrum, Fig.~2.  The less successful reconstructions, which
also yield a higher $d_L$ are shown in the Table, give a FS in which the first
peak is very broadened.

Many spurious detections of chaos have been made in the literature, especially
when techniques such as correlation dimensions are used, and one should
therefore always be wary of such pitfalls.  SKB and BKSM did some tests that
concluded that it is difficult to fool the global flow reconstruction method
into erroneously predicting low-dimensional chaos on a stochastic signal.
(More specifically they had applied the method to an average periodic component
of the R~Sct lightcurve and contaminated with colored noise of same FS
as R~Sct).

As a test for the stability of the reconstruction for AC~Her we perform
the following two experiments:

 The first one is to take a section of our best chaotic synthetic signal (shown
in Fig.~5 (second row) and Fig.~6 (Col.~5)  of the same length and sampled as
the AAVSO data set, add Gaussian noise with the observational intensity of
0.15~mag, perform a 3~day averaging and a cubic spline smoothing with
$\sigma$=0.065.  The BK projections of this signal are shown in col.~6 of
Fig.~6.  When this signal is then subjected to the global flow reconstruction:
Interestingly, the reconstructed synthetic signal is again chaotic.  Its
Lyapunov dimension is found to be $d_L\approx$ 2.2--2.4, i.e. very close 
to the values for AC~Her lightcurve. It is nevertheless remarkable that 
this second processing of the signal still yields chaos with similar 
properties as the first one.
 
The second test signal is taken to be stochastic.  Starting with the periodic,
8~harmonic signal used to prewhiten the AC~Her data in Fig.~2a, we added noise
with the observational intensity of 0.15~mag.  It is gratifying that just as in
BKSM, {\sl no} erroneous chaotic synthetic signals are produced by the
reconstruction for any of the values of $d_E$, $\Delta$, $p$, $\sigma$ that we
have tried.

\titlea{Discussion}

In \S2 we showed that the lightcurve of AC~Her is not compatible with an
interpretation of a steady periodic or multiperiodic star, nor with a
multiperiodic and evolving one.  The next simplest working hypothesis is that
the lightcurve is chaotic, i.e. that it is generated by a low dimensional
dynamics with chaotic trajectories.  In \S3 we show that indeed we can
explicitly construct dimensional flows with simple polynomial nonlinearities
with properties close to the observed AC~Her lightcurve. Furthermore, whenever
the synthetic solutions resemble the observational data, they have similar
Lyapunov exponents and Lyapunov dimension, this independently of reconstruction
dimension $d_E$.

The reconstruction of maps is less robust than it is for R~Sct (BKSM).  We see
several reasons for this.

 \ni(a) First, the signal to noise ratio is lower, and the observational error
is a little larger, viz. 0.15 as compared to 0.10 for R~Sct, because AC~Her is
fainter.  The reconstruction method therefore has greater difficulty
reconstructing the attractor from the noise.

 \ni(b) The AC~Her pulsations are more regular, and the trajectory explores a
smaller region of phase space.  Especially there is a complete absence of
trajectories about the unstable fixed point of the map (or flow).  This also
has the drawback of determining less well the linear part of the map and the
properties of the fixed point; In BKSM we linearized the map to show that its
unstable fixed point has 2 spiral roots, one unstable and one stable with
the trajectory running away from the fixed point with the fundamental frequency
to return toward the fixed point with approximately twice the frequency.  This
was interpreted as showing that the chaotic behavior of R~Sct is the result of
the nonlinear interaction of two vibrational modes that are in an approximate
resonance.  A similar argument is thus not possible here.

 \ni (c) The more irregular features in the observational data are too few to
have a sufficient influence on the map construction, e.g. the peak near 220~d
and the 'unusual' oscillation near 3500~d.  These features are also prominent
as the few very large excursions in the BK plots of the first column of Fig.~6.

When these wildest excursions are disregarded the AC~Her projections and the
$\Delta$=6 reconstructions become quite similar.  This is quite apparent in
the noisy smoother synthetic signal of Col.~6 of Fig.~6 that has already been
discussed.

\vskip 6pt

From the inequality $d_L < d \leq d_E$ and the value $d_L\approx 2.2$ that
we have determined we conclude (cf. SKB) that the physical Euclidean
dimension of the underlying attractor is therefore 4, and perhaps as low
as 3.

\titlea{\ \ Conclusions}

\jumpover{
 That only two modes should be involved is not astonishing on theoretical
grounds for two reasons. First, most radial pulsators involve only a few modes
of oscillation (for a review \cf \eg Buchler 1993).  A necessary condition
for the occurrence of irregular pulsations is that the amplitude modulations
occur on the same time scale as the pulsations themselves, in order words, the
relative growth rates must be of order unity.  Linear stability analyses of
models show that this is indeed the case for W Vir and even more so for their
longer period siblings, the RV Tau stars.  (In contrast, in the case of the
classical variable stars the relative growth or decay rates of the dominant
modes are very small, at most of the order of a few percent for Cepheids, for
example. The pulsations of these stars are therefore regular, \ie periodic or
multiperiodic).  Second, a Floquet analysis of the model pulsations of W Vir
stars showed that the onset of the period doubling cascade in these models is
associated with the 5:2 resonance between the fundamental mode and the second
overtone (Moskalik \& Buchler 199?). The strongly dissipative (or
'nonadiabatic') nature of these pulsations is expected to broaden considerably
the resonance region.  It is therefore likely that a similar two-mode
interaction persists for the more luminous RV Tau stars.
 }

AC~Her is the second irregular variable star that we have analyzed with the
global flow reconstruction method.  Our conclusion is that the pulsations of
AC~Her are very likely to be the result of low dimensional chaos.  However,
because of the inferior quality of the observational data the case is not
quite as strong as for R~Sct for which we have little doubt as to the chaotic
nature of the pulsations.  We also recall that the numerical hydrodynamical
modelling of the pulsations of W~Vir type stars (KB88 and Serre, Koll\'ath \&
Buchler 1995b) showed very clearly low dimensional chaos.

The applications of nonlinear analyses such as this one are still in their
infancies.  Furthermore, the available observational data are far from optimal
for such applications.  Further progress will have to await more specific
observing efforts tailored to the type of data that are required by nonlinear
analyses, such as long uninterrupted observational spans.  We stress however
that long seasonal gaps do not pose a problem for nonlinear analyses.

When the results of our analysis of AC~Her are compared with those of R~Sct, on
the one hand, and with the pulsations of W~Vir models one notices a trend,
which is already obvious from a visual inspection of he lightcurves: As the
period increases the alternations in the shallow and deep minima as well as the
modulations become increasingly irregular.  This behavior is also observed in a
quantitative fashion in the Lyapunov dimension that increases from $\approx$
2.05 for the W~Vir model, to $\approx$ 2.2 for AC~Her and to $\approx$ 3.1 for
R~Sct.  It appears that one may be able to extract useful quantitative
information from an irregular lightcurves.  In turn, that this will put novel
and very useful constraints on the numerical modelling of these objects.  On
the theoretical side a thorough numerical hydrodynamical survey of W~Vir and
RV~Tau models is clearly also necessary to confirm this hope.

\titlea{\ \ Acknowledgments}

This research has been supported in part by NSF (AST92-18068, AST95-28338,
INT94-15868), a Hungarian OTKA grant (F4352), an RDA grant at UF, the French
Minist\`ere pour la Recherche et l'Espace, and RCI grant from IBM through UF.

\begref{References}

\ref 
 Abarbanel, H. D. I., Brown, R., Sidorowich, J. J., Tsimring, L. S. 1993,
Rev. Mod. Phys. 65, 1331

\ref %ok
 Buchler, J. R., 1997, {\sl Search for Low-Dimensional Chaos in Observational
Data}, International School of Physics "Enrico Fermi", Course CXXXIII on {\sl
"Past and Present Variability of the Solar-Terrestrial System: Measurement,
Data Analysis and Theoretical Models"}, Eds. G. Cini Castagnoli \&
A. Provenzale (in press).

\ref %ok
 Buchler, J.R. 1993, in {\sl Nonlinear Phenomena in Stellar Variability}, p.~9,
Eds. M. Takeuti \& J.R. Buchler, Dordrecht: Kluwer Publishers,
reprinted from 1993, Ap\&SS, 210

\ref %ok
 Buchler, J. R., Serre, T., Koll\'ath, Z. \& Mattei, J. 1995, Phys. Rev.
Lett. 74, 842 [BSKM]

\ref %ok
 Buchler J. R., Koll\'ath, Z., Serre, T. \& Mattei, J. 1996, ApJ 462, 489
[BKSM]

\ref %ok
 Gouesbet, G., LeSceller, L, Letellier, C., Brown, R., Buchler, J.R. \&
Koll\'ath, Z. 1997, {\sl Nonlinear Signal and Image Processing},
Ann. N.Y. Acad. Sci. Vol.~808, p.~25.

\ref %ok
 Kov\'acs, G. \& Buchler, J. R. 1988, ApJ, 334, 971, [KB88]. % reg + irreg

\ref 
 Ott, E. 1993, {\it Chaos in Dynamical Systems} (Univ. Press: Cambridge) 

\ref
 Percy, J.R., Bezuhly, M. Milanowski, M. \& Zsoldos, E. 1997.
 PASP 109, 264

\ref
 Percy, J.R. \& Mattei, J.A.  1993. Ap\&SS 210, 137
 PASP 109, 264

\ref %ok
Serre, T., Koll\'ath, Z. \& Buchler, J. R. 1995a, A\&A 311, 833 [SKB]

\ref %ok
Serre, T., Koll\'ath, Z. \& Buchler, J. R. 1995b, A\&A 311, 845

\ref
Weigend, A. S. \& Gershenfeld, N. A. 1994, {\it Time Series Prediction}
 (Reading: Addison-Wesley)

\ref 
Zsoldos, E. 1988, IBVS, 3192

\end\bye